\documentclass[submission,copyright]{eptcs}
\providecommand{\event}{Wivace 2013} 
\usepackage{breakurl}             

\usepackage{color}
\usepackage{ulem}
\usepackage{epsfig}
\usepackage{graphicx}
\usepackage{epstopdf}
\usepackage{amsmath}
\usepackage{amsfonts}
\usepackage{bm}
\usepackage{bbm}
\usepackage{amssymb}
\usepackage{mathrsfs}
\usepackage{color}
\usepackage{tabularx}
\usepackage{chemarrow}

\title{Recent developments in research on catalytic reaction networks}
\author{Chiara Damiani
\institute{Dept. of Informatics, Systems and Communication,\\ University of Milan Bicocca, Italy}
\email{chiara.damiani@unimib.it}
\and
Alessandro Filisetti 
\institute{European Centre for Living Technology,\\ University Ca' Foscari of Venice, Italy }
\email{alessandro.filisetti@unive.it}
\and
Alex Graudenzi 
\institute{Dept. of Informatics, Systems and Communication,\\ University of Milan Bicocca, Italy}
\email{alex.graudenzi@unimib.it}
\and
Marco Villani and Roberto Serra
\institute{Dept. of Physics, Informatics and Mathematics, Modena and Reggio Emilia University}
\email{marco.villani@unimore.it \quad\qquad rserra@unimore.it}
}

\def\titlerunning{Catalytic Reaction Nets research}
\def\authorrunning{Damiani et al.}
\begin{document}
\maketitle

\begin{abstract}Over the last years, analyses performed on a stochastic model of catalytic reaction networks have provided some indications about the reasons why wet-lab experiments hardly ever comply with the phase transition typically predicted by theoretical models with regard to the emergence of collectively self-replicating sets of molecule (also defined as autocatalytic sets, ACSs), a phenomenon that is often observed in nature and that is supposed to have played a major role in the emergence of the primitive forms of life.\\
The model at issue has allowed to reveal that the emerging ACSs are characterized by a general dynamical fragility, which might explain the difficulty to observe them in wet-lab experiments.\\
In this work, the main results of the various analyses are reviewed, with particular regard to the factors able to affect the generic properties of catalytic reactions networks, for what concerns not only the probability of ACSs to be observed, but also the overall activity of the system, in terms of production of new species, reactions and matter.

\end{abstract}

\section{Introduction}
\label{introduction}
Cells are the basic unit of life and one great challenge shared by different fields, such as synthetic biology, systems chemistry and origin of life, is to find the appropriate conditions under which they can effectively emerge, persist and evolve.\\
Cell replication and adaptation emerge from the coordinated cooperative dynamics of several biochemical networks, which are enclosed into a compartment. A metabolism provides the resources required to build and maintain cellular components and allows the cell to grow until eventually splitting into two distinct copies. In order for cells to adapt to a changing environment, information carriers are required as well, and they must be coupled with metabolic reactions.\\
In regard to this, the presence of one or more auto-catalytic sets (ACS) of molecules inside a reactions network may be a fundamental property to achieve a certain degree of robustness, leading to the formation of self-maintaining structures able to self-replicate. 

Several theoretical models~\cite{Dyson:1985uq,Eigen1977a,Eigen1988,Eigen1978,ac:ES78c,Jain:1998fk,Kaneko:2006eu,Kauffman:1986mi,Segre2000,Segre:1998sf} have therefore been proposed to support the experimentalists in addressing this issue. In spite of their intrinsic differences, these models typically predict a phase transition such that, once certain conditions are met (generally relative to the level of heterogeneity of the chemical soup), ACSs spontaneously emerge. Nevertheless, collective self-replication has hardly ever been observed in wet-lab. \\
The stochastic model of catalytic reaction networks introduced in~\cite{Filisetti:2010fk} has proven appropriate for investigating the rationale at the basis of this mismatch between theoretical predictions and experimental observations.

The model takes its first steps from the association of a temporal dynamics to Kuaffman's model~\cite{Kauffman:1986mi}, according to which every molecule has a certain probability to catalyze a reaction,  and it is therefore without knowledge of which molecules need to be present in the mixture, and without knowledge of which specific molecule is catalyzing which reaction.\\ 
The main credit of the model is that, by treating the dynamics as stochastic, it allows to take into account the effect of noise, random fluctuations and low-numbers variations. In fact, studies performed on this model~\cite{JSC2011,Filisetti2011b,TIB2011} have revealed that, in most cases, the emergence of ACSs critically depends upon the existence of some rare molecules, which may disappear as a result of fluctuations. This outcome has provided a possible explanation of the failure in observing ACSs in wet lab experiments.

Furthermore, the model has revealed well suited to investigate the conditions that affect not only the probability of ACSs to emerge, but the overall dynamical properties of catalytic reaction networks.
At this aim, the aforementioned studies have analyzed different ensembles of random catalytic reaction networks, where the ensembles are characterized by different structural constraints relative for example to the environmental conditions (e.g. influx diversity, residence time) or to the introduction of energy transport and utilization dynamics. Members of the same ensemble differ one another in terms of \textit{chemistry} of the system, which will be defined in the following.\\
The idea moves from Kauffman hypothesis~\cite{ensemble} that life had, as its initial substrate, the behavior of randomly aggregated reaction nets. Therefore, the average properties of the ensemble members may be able to provide insight into the structure and dynamics of real networks.

This paper is intended to review the major results of these analyses, with particular regard to those conditions that affect the overall activity of the system and the emergence of ACSs. In Section~\ref{model} the main features of the model are recalled, in Section~\ref{results} the descriptors of the dynamics are illustrated and their correlation with each of the factors that have been considered is presented. Section~\ref{robustness} is dedicated to a general remark regarding the fragility of the observed ACS. Section~\ref{conclusions} draws some conclusions and perspectives.
%
\section{The model}
\label{model}
\subsection{Main features of the model}
\label{sect:the_model}
\subsubsection*{Entities and interactions}
The system is modeled as a set of simple molecules within an open system, which are represented as a sequences of letters  (or  {\em bricks}), oriented from left to right, taken from a finite alphabet (e.g. $A$ and $B$ are the alphabet considered in this work) and can be either monomer (one letter) or polymers.
Molecules that are characterized by an identical sequence are considered as multiple copies of the same species $x_i, i=1,2,...,N$, $X = [x_1, x_2, \dots, x_N]$.

\begin{figure}[ht]
\begin{center}
\includegraphics[width=8cm]{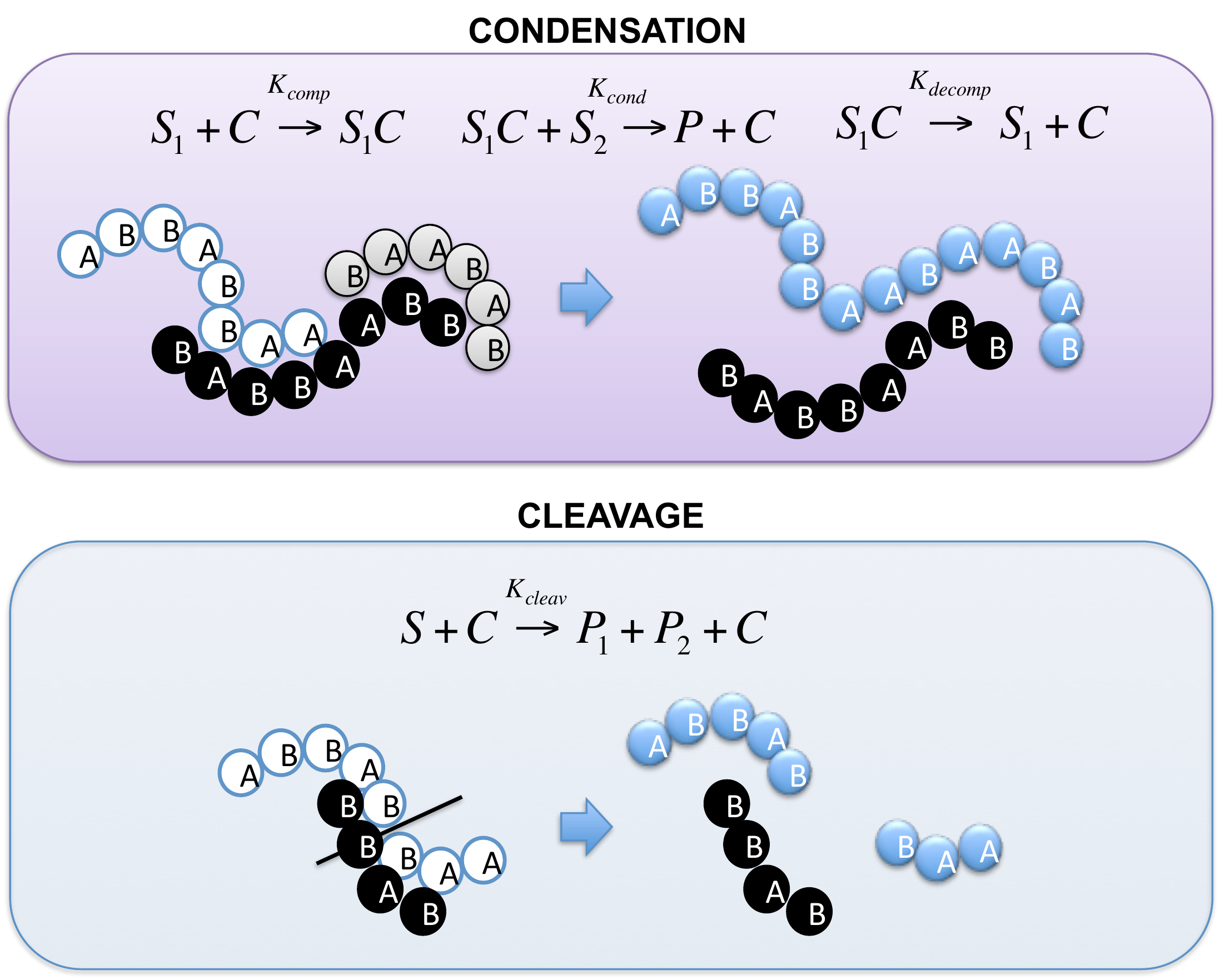}
\label{schemino}
\caption{A graphical representation of the two types of reactions. TOP: Condensation reaction. The white and the grey chains are two substrates, the black chain is the catalyst and the blue chain the product. The reaction occurs in two steps, the binding of a temporary molecular complex and the formation of the product. BOTTOM: Cleavage reaction. The white chain is the substrate, the black chain the catalyst and the blue chains are the two products of the reaction. }
\end{center}
\end{figure}

Different species are allowed to interact with each other according to two basic reactions:  {\em cleavage}, which cuts a species into two shorter species~\footnote{The species that is cut must be composed of more than one brick.} and {\em condensation}, which concatenates two species into a longer one. It should be noted that both reactions require the participation of a third species, which plays the role of a catalyst, in order to occur. Hence, we exclude the presence of spontaneous reactions by assuming a sufficiently high activation energy for each reaction. A template of the two reaction types is given in Figure~\ref{schemino}, along with a graphical example.\\
It can be noticed that condensation is a third order reaction involving two substrates and a catalyst. As catalysts generally bind to one of the substrates first, forming a temporary complex substrate-catalyst, condensation is modeled as two separate reactions: the former takes the shape  $S_1 + C  \xrightarrow{K_{comp}} S_1\Join C$ and the latter  $ S_1\Join C + S_2  \xrightarrow{K_{cond}} P + C$; where $C$ is the catalyst, $S_1$ and $S_2$ are the two substrates, $K_{comp}$ and $K_{cond}$ are the kinetic rates of the two reactions respectively and $P$ is the final product, that is  the final condensation between the two substrates. As an exception to the assumption made above, the temporary complex is let dissociate spontaneously at a small kinetic rate $K_{decomp}$ according to the reaction  $S_1\Join C  \xrightarrow{K_{decomp}} S_1 + C$.\\
On the contrary, cleavage is a second order reaction and is modeled as a single reaction:  $S + C \xrightarrow{K_{cleav}} P_1+ P_2 + C$, where $P_1$ and $P_2$ are the products originating from substrate S and $K_{cleav}$ is the kinetic rate at which the cleavage reaction occurs. \\
Within this work, as a first approximation, we consider that the rate $K_{cleav}$ takes the same value for all cleavage reactions, the rate $ K_{comp}$ takes the same value for all complex formation reactions and so on for $K_{decomp}$ and $K_{cond}$~\footnote{In particular $K_{cleav}$ is set to 25 $M^{-1}sec^{-1}$ for all the cleavage reactions,  $K_{comp}$ is set to 50 $M^{-1}sec^{-1}$ for all the complex formation reactions, $K_{decomp}$ is set to 1e-06 $sec^{-1}$ for all the complex dissociation reactions and $K_{cond}$ is set to 50 $M^{-1}sec^{-1}$ for all the end condensation reactions. The choice of these values has been partially arbitrary; a sensitivity analysis regarding these parameters can be found in~\cite{sensitivityChiara}.}.
It is worth noticing that, exception made for the complex dissociation reaction, backward reactions are neglected, we indeed hypothesize that the Gibbs energy $\Delta G$ for any reaction is sufficiently large. 

Once the form of interaction between species has been defined, the actual players involved in each possible reaction are assigned at random, giving rise to what we call the $chemistry$ of the system, with the key restriction that only species that posses a sufficient ''level of complexity'' are allowed to be catalyst for a reaction. We suppose that the level of complexity depends on the number of bricks composing a species, e.g. within this work monomers and dimers cannot be catalyst of any reaction.
%
In this regard, the number of possible reactions, over the number of conceivable ones, depends on the reaction probability $r$ as explained below. \\
Any species $x_i$ has a finite probability $r_i$ of being chosen as catalyst of a randomly chosen reaction among those belonging to the set $R$ of all conceivable reactions. Because, each species in the system could condensate with any other species in the system, or be split at any cutting point into smaller species, the cardinality of the set R of all the conceivable reactions for the system, at a given time, is therefore given by:
\begin{equation}
R=\sum_{i=1}^{N}(L_{x_{i}}-1)+N^{2}
\label{eq:conceivableRcts}
\end{equation}
where $N$ is the cardinality of $X$ and $L_{x_{i}}$ is the length of $x_i$ (i.e. the number of bricks of that specific species). The first term of Eq.~(\ref{eq:conceivableRcts}) refers to the conceivable cleavages and the second one to the conceivable condensations.\\
We remark that, although the reaction network is built probabilistically, it remains invariant in time; in other words, once a species is chosen to be the catalyst of a given reaction, that species will always be catalyst for that reaction. Nevertheless, the reaction network may expand in time. Indeed, because new species are allowed to form during the evolution of the system, the cardinality of R may change and the reaction network must be updated to include the new species, provided that previously existent chemistry is kept unaltered. 

The model reactions are supposed to occur within a continuously stirred open-flow tank reactor (CSTR). Therefore, together with the chemistry of the system, the $initial~set$ and the composition of the $incoming~flux$ must be defined. The initial set, which is the population of molecules that exist within reactor when the simulation starts, is selected by drawing a uniform number of molecules, according to the desired overall concentration, for each of the species up to a given length $L_{max}$. Along similar lines, the $incoming~flux$, that is, the molecules that continuously enter the system, depend on the desired amount of molecules per time entering the system and on the maximum length of the species that are allowed to enter it.  
The outgoing flux is instead determined only by the average $residence~time$ of each molecule in the system, i.e. the outgoing flux is composed of all the molecules in accordance with their concentrations. 

\subsubsection*{Energy carriers}
\label{EC}
An extension of the original model described in the previous section takes into account also the role of energetic constraints, by introducing some molecules, hereinafter $Energy~Carriers$ (EC), able to storage energy as chemical bond, which are in charge of carrying energy to those reactions that necessitate it in order to occur.

The reactions composing the chemistry of the model are therefore subdivided into $exergonic$ or $endoergonic$ depending on their energetic constraints. The latter, as opposed to the former, may occur if and only if some of the elements involved in the reaction have obtained an amount of energy from a carrier.\\
An extra type of reaction is therefore added to the aforementioned conceivable reactions (Figure~\ref{schemino}) and regards the binding of a substrate to the energy carrier as follows: $S + EC  \xrightarrow{K_{nrg}} S+$, where $K_{nrg}$ is the reaction rate constant, also reffered to as \textit{energization kinetic constant}, and $S+$ is the energized form of the species under issue. It should be noticed that, once the EC has released the energy, it is removed from the system, whereas the species remains energized until it participates to an endoergonic reaction.\\
As better elucidated in~\cite{Filisetti2011b}, which molecule or molecules among the players of a given reaction must carry the energetic group in order for the reaction to occur (e.g. the catalyst, some of the substrates or both), can be decided according to a Boolean Function.

So far, the analyses of the model has been limited to the case in which at least one of the substrates of the reaction, but not the catalyst, must be energized and in which all condensation reactions are considered as endoergonic, whereas all cleavage reactions are not.
\subsection{Model dynamics}
\subsubsection*{The simulation algorithm}
The system's dynamics is simulated by means of an extension of the well-known Gillespie algorithm~\cite{Gillespie:1977fv} for the stochastic simulation of chemical reaction systems made to allow the generation of novel species and novel reactions. In fact, cleavage and condensation of elements either initially present within the reactor or entering it from the external environment can generate new species, which, in turn, can be involved in new reactions, both as catalysts and substrates, thus requiring the variable space to increase in size and complexity. 

Moreover, in order to speed up the computational time we perform hybrid simulations by modeling some processes, such as in-flux and out-flux ones, as continuous rather than stochastic. In particular, these are modeled as ODEs, where  the time interval $\Delta t$ between two successive reactions is let be determined by the Gillespie algorithm, whereas the the actual species to be either introduced or washed out are chosen probabilistically in accordance to their relative concentrations, as better explained in~\cite{JSC2011}. The processes related to i) energy carries influx; ii) species energization and de-energization; iii) energy carriers decay; are modeled according to this method as well.
\subsubsection*{ACS detection}
\label{ACS}
As the simulation time proceeds, different reactions of the chemistry can occur at different frequencies. By observing which reactions have occurred within a temporal window, one can look for the presence of autocatalytic sets.

It is worth mentioning that an ACS may be defined in alternative ways: following~\cite{Kauffman:1986mi}, we consider an ACS as a subset of chemicals whose production is catalyzed by at least one other member of the subset. 

According to this definition, a graph deciphering the catalytic activity of the system reveals useful in order for ACSs to be detected: each node in the graph is a species of the model, and an edge between two species $A$ and species $B$ is drawn if $A$ has catalyzed the production of $B$ at least once within the temporal window. 
Once the graph is depicted, the presence of an ACS, within the temporal window, can be easily detected via strongly connected component analysis or via the study of the eigenvalue with the largest real part (the reader is referred to~\cite{TIB2011,Hordijk2010,Hordijk:2004vl,Mossel2005}).

It should be noted that, in order to avoid considering trivial cases, we prevent the presence of ACSs within the incoming flux; in this way, we force the formation of ACSs to require species which come to exist within the reactor.\\
\section{Conditions affecting the system dynamics}
\label{results}
\subsection{Dynamics descriptors}
\label{variables}
As already mentioned in Section~\ref{introduction}, the investigation of the generic properties of catalytic reaction networks, with specific regard to the emergence of autocatalytic sets, is a major issue in systems chemistry and, particularly, in the research concerning the origin of life. The stochastic model illustrated in Section~\ref{model} has demonstrated well suited to investigate the conditions that favor the emergence of ACSs. Because the presence of autocatalytic cycles may lead to a significant departure of the concentrations of the elements belonging to the cycle from the expected one, analyses of the model dynamics, besides considering some indicators strictly connected with the emergence of ACS, have paid attention to the concentration of the species that resides in the CSTR~\footnote{It should be noted that the overall concentration of species within the CSTR tends to an equilibrium level, regardless of external conditions.}, with particular regard to those of new species (species that are not already present either in the initial set or in the incoming flux), and to the number of such species. 

The following variable of the system have therefore been taken into account:
\begin{itemize}
\item Number of ACSs in time, where the time is further divided into a given temporal window and the presence of ACSs is detected into any temporal window~\footnote{The results described in this paper typically refer to a temporal window of 10 sec.} (see Section~\ref{ACS})
\item Number of species into an ACS in time: how many species are involved in the detected ACSs 
\item Probability of an ACS to be observed in time: in how many chemistries, or simulations of the same chemistry, at least an ACS is observed
\item Number of molecules in ACS in time: amount of molecules of the species involved in an ACS
\item Number of new species in time: how many species are present (with at least one molecule) within the CSTR that are not also part of the initial set or incoming flux
\item Number of molecules (not influx) in time: how many molecules are present within the CSTR that belong to species that are not also part of the initial set or incoming flux
\end{itemize}

Besides the obvious phenomenon according to which the emergence of ACSs is associated with an increase in the average connectivity of the chemistry and it is described by a transition phase (as illustrated in ~\cite{JSC2011}), other conditions (or parameters) of the system are likely to affect the variables listed above, in particular: the composition of the incoming flux, the residence time and the availability of energy. The variables of interest have therefore been analyzed in relation to variations in the parameters relative to  these aspects, as better detailed in the following subsections, whereas the probability $r$ that a given species catalyses a certain reaction is maintained at the critical value $1.03x10{-3}$,  that is the value at which the transition between non-emergence and emergence of ACSs occurs (following ~\cite{JSC2011}), for all the species in the system, $r_i=r, i=1,2,...,N$.\\ 

The relation between parameters and variables is generally assessed by sampling a bunch of different chemistries and by simulating different stochastic realizations of each chemistry~\footnote{It is worth mentioning that, because during a single simulation new species may emerge, the reaction scheme must be updated for all the different stochastic simulations of the same chemistry.} for 1000 sec., for different values of the parameter under study. Then, for each parameter value, the variable of interest is averaged over all the simulations at a given time point, and the average value of the variable is analyzed as a function of the parameter and of time. 

The correlations between parameters and variables that have been identified are schematically summarized in Table~\ref{tabella}.
\begin{table}
\caption{The table summarizes the outcomes of the analyses about the correlation between parameters and descriptors of the dynamics. Each row is a different parameter, the rows are grouped in families of parameters relating to a common aspect. The symbol next to the parameter name refers to its variation (in this case it always is a positive variation). Each column refers to one of the descriptor dynamics listed in Section~\ref{variables}. The symbol at the interception between rows and columns refers to the variation observed in the corresponding dynamics descriptor as a function of the parameter at issue (the other parameters being invariant):  \~{}  stands for a non-significative variation; $\uparrow$ stands for a positive variation of the variable; $\downarrow$ stands for a negative variation; whereas $\nearrow \searrow$ represents the fact that an initial positive variation is followed by a negative one, implying therefore the existence of an optimal value. Empty cells refer to missing information.}
\label{tabella}
\centering
\includegraphics[width= 12cm]{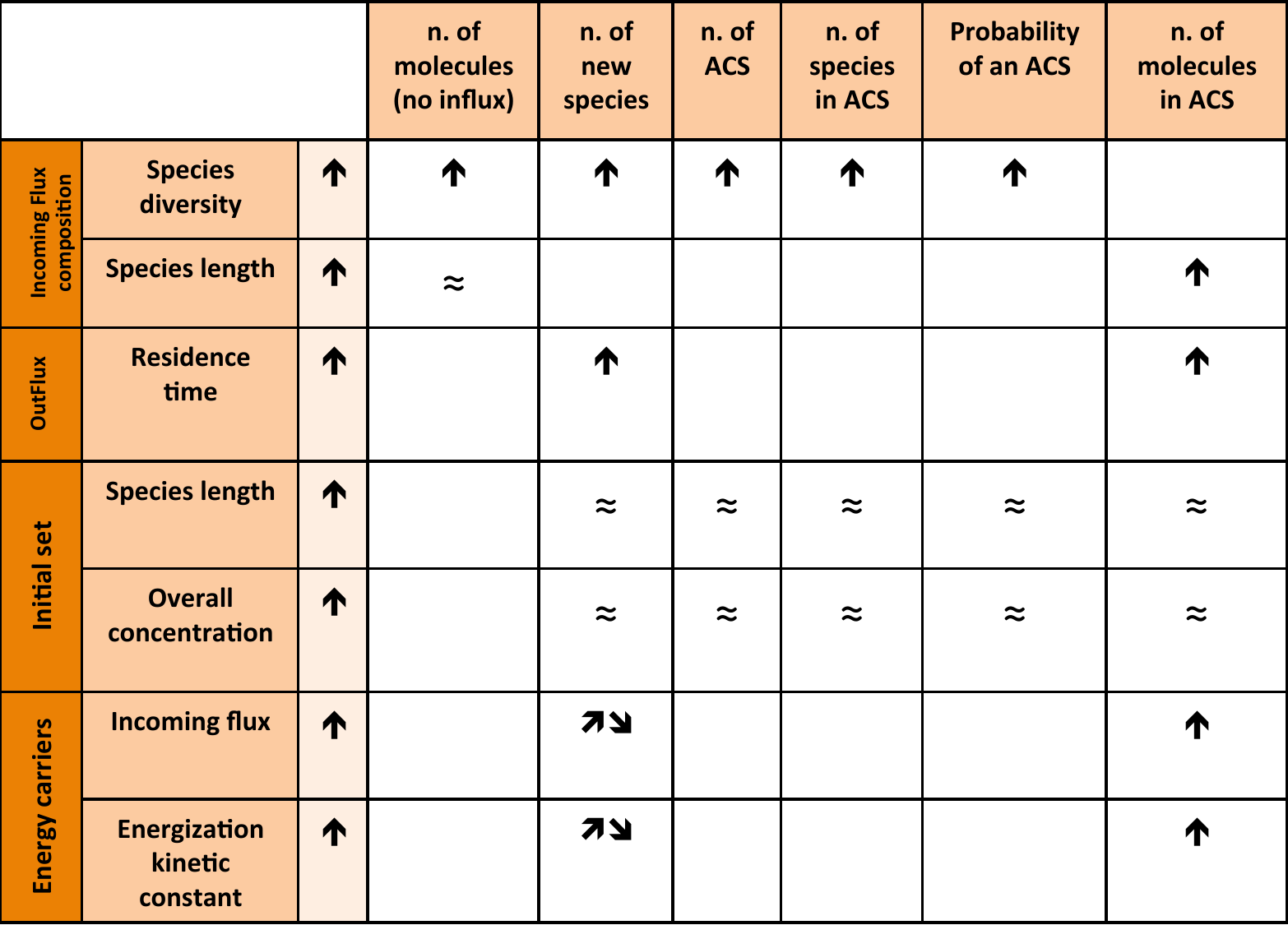}
\end{table}
\subsection{Influence of incoming flux}
The incoming flux is characterized by how many molecules enter the CSTR per unit of time and the species they belong to. The influx can be composed of all the species up to a maximum length $M_{in}$ or of a subset of such species, according to  a probability distribution. It is therefore reasonable to expect both the typical length of the influx species and the variety of the species entering the CSTR to correlate with the system dynamics.

The analyses described in~\cite{JSC2011} and summarized in Table~\ref{tabella} have however revealed that, if on the one side the number of different species deeply affects the dynamics of the system in all respects, on the other side the system seems rather insensitive to variations in their typical length.\\
%
%
\subsection{Influence of residence time}
Given that all the molecules within the CSTR are allowed to leave the system with rate $K_{out}$, the inverse of $K_{out}$ expresses the average residence time of a molecule within the CSTR. This parameter is supposed to have some effects on the probability of a given molecule to react with others. It is therefore meaningful to investigate the dependency of the variables of interest listed above on this parameter.

In order to isolate the effect of the residence time from that of a reduced presence of molecules within the reactor, in~\cite{TIB2011} a proper combination of the values of $K_{out}$ and $K_{in}$ (the rate of the incoming flux) has been set in order to keep the number of molecules present in the system at the equilibrium, while varying the residence time.

The study has demonstrated that the residence time has indeed repercussions on the systems dynamics, with larger residence times entailing a general enhancement of the activity of the system: the creation of new species is favored and a higher amount of molecules is involved into ACSs.

\subsection{Influence of initial set}
As well as the molecules entering the systems through the incoming flux, also the molecules that are already present within the reactor when the simulation starts may play a role in the probability of creating new species and give rise to ACSs. Nevertheless, in~\cite{TIB2011}, it has been shown that neither a variation in the maximum allowed length of the initial species nor in their relative abundance in terms of initial concentration have a significant impact on the creation and maintenance of new species and on the probability of ACSs to emergence.

\subsection{Influence of energy transport and utilization dynamics}
The introduction of energy carriers in the model, mentioned in Section~\ref{EC}, allows to investigate the influence of a variation in the energy carriers incoming flux rate $\phi E$ or in the energization kinetic constant $k_{nrg}$ on the overall dynamics.

The preliminary analyses presented in~\cite{Filisetti2011b} have shown that the average number of species (not belonging to the incoming flux) present in the reactor at the end of the simulation as a function of $\phi E$ and of $k_{nrg}$ exhibits a maximum portion of surface after which the variable begin to decrease, pointing at an optimal amount of energy for the system in terms of overall production of new species~\footnote{It should be noticed that the effect is partially due to the the particular assumptions concerning the chosen energy function.}.\\
The presence of an optimal level of available energy for the enhancement of the general activity is confirmed also by a unique maximum peak in the same region in the number of molecules produced within ACSs.

\section{ACSs Robustness}
\label{robustness}
An important remark concerns the degree of robustness of the ACSs that have been observed across all the conditions analyzed in the previous Section.

In~\cite{JSC2011,TIB2011,Filisetti2011b} it has been shown that, in general, the concentration of the species belonging to the identified ACSs do not exhibit a significant departure from the value they would have if they wouldn't be part of an ACS. This consideration implies that the observed ACSs do not comply with the definition of Autocatalytic Metabolism given by Bagley et al.~\cite{Bagley1989}.
Moreover, in most of them, the catalytic closure is achieved by means of a ÒbottleneckÓ reaction that occurs rarely during the temporal window, in some cases even only once. This makes an ACS fragile, as the probability that in the next temporal window that specific reaction will occur is almost negligible.

It is worth stressing that, although a higher diversity of the influx (in term of species), a longer residence time and the presence of energy constraints, are all factors that favor the emergence of structural ACSs, they do not confer a major robustness to them, and no form of self-sustaining dynamics is observed.

\section{Conclusions and further remarks}
\label{conclusions}
This review has provided a global perspective on the outcomes of all the analyses performed on the stochastic model of catalytic networks introduced in~\cite{Filisetti:2010fk}, which are aimed at filling the gap between theoretical predictions and experimental findings, in regard to the emergence of those autocatalytic sets of molecules that are so widespread in nature.

It has been observed that either a variation in some parameters or a release in the model assumptions (i.e. the extension of the model in order to take into account energization dynamics) have the power to increase the probability to observe an ACS, as well as the number of species and molecules involved in it, and to foster the generation and maintenance of new species within the reactor. In spite of this, a general conclusion is that the emergence of an ACS does not entail a significant take off of the concentration of the species belonging to it.

The reason for this mismatch lays in the fact that these ACSs are fragile, in the sense that some of the edges of the strongly connected component refer to reactions that occur only a few time within the considered temporal window. Because the kinetic rates are identical for all condensation and cleavage reactions, this phenomenon is necessarily due to a low concentration of the species at issue. A low concentration of a species, within the stochastic framework, is likely to result in its complete extinction, leading to the destruction of the ACS itself, hence its fragility.
It is remarkable that this property of ACSs would be rather difficult to detect within a deterministic framework, which do not straightforwardly accounts for species extinction (or creation).

A remarkable hypothesis that can be inferred from these results is that the assumption that reactions take place within a chemostat, i.e. a continuously stirred open-flow tank reactor, might be too restrictive, pointing at a major role played by the presence of a compartment similar to that of membrane in cells for robust ACSs to emerge from a catalytic reaction network.\\
The investigation of this hypothesis is of great scientific interest and it is worth been addressed with a novel model of protocell coupled with the dynamics of stochastic catalytic reaction networks.

\section{Acknowledgments}
The research leading to this paper has been partially funded by the EU funded project INSITE GA n¡ 271574 under the FP7.

\bibliographystyle{eptcs}

\bibliography{personalBib,sensitivity,refsFinal}

\begin{thebibliography}{10}
\providecommand{\bibitemdeclare}[2]{}
\providecommand{\surnamestart}{}
\providecommand{\surnameend}{}
\providecommand{\urlprefix}{Available at }
\providecommand{\url}[1]{\texttt{#1}}
\providecommand{\href}[2]{\texttt{#2}}
\providecommand{\urlalt}[2]{\href{#1}{#2}}
\providecommand{\doi}[1]{doi:\urlalt{http://dx.doi.org/#1}{#1}}
\providecommand{\bibinfo}[2]{#2}

\bibitemdeclare{article}{Bagley1989}
\bibitem{Bagley1989}
\bibinfo{author}{R~J \surnamestart Bagley\surnameend}, \bibinfo{author}{J~D
  \surnamestart Farmer\surnameend}, \bibinfo{author}{S~A \surnamestart
  Kauffman\surnameend}, \bibinfo{author}{N~H \surnamestart Packard\surnameend},
  \bibinfo{author}{A~S \surnamestart Perelson\surnameend} \&
  \bibinfo{author}{I~M \surnamestart Stadnyk\surnameend}
  (\bibinfo{year}{1989}): \bibinfo{title}{{Modeling adaptive biological
  systems.}}
\newblock {\sl \bibinfo{journal}{Bio Systems}}
  \bibinfo{volume}{23}(\bibinfo{number}{2-3}), pp. \bibinfo{pages}{113--37;
  discussion 138}.
\doi{10.1016/0303-2647(89)90016-6}
\newblock \urlprefix\url{http://www.ncbi.nlm.nih.gov/pubmed/2627562}.

\bibitemdeclare{article}{sensitivityChiara}
\bibitem{sensitivityChiara}
\bibinfo{author}{Chiara \surnamestart Damiani\surnameend},
  \bibinfo{author}{Alessandro \surnamestart Filisetti\surnameend},
  \bibinfo{author}{Alex \surnamestart Graudenzi\surnameend} \&
  \bibinfo{author}{Paola \surnamestart Lecca\surnameend}
  (\bibinfo{year}{2013}): \bibinfo{title}{Parameter sensitivity analysis
  of stochastic models: Application to catalytic reaction networks}.
\newblock {\sl \bibinfo{journal}{Computational Biology and Chemistry}}
  \bibinfo{volume}{42}(\bibinfo{number}{0}), pp. \bibinfo{pages}{5 -- 17},
  \doi{10.1016/j.compbiolchem.2012.10.007}.
\newblock
  \urlprefix\url{http://www.sciencedirect.com/science/article/pii/S1476927112000771}.

\bibitemdeclare{book}{Dyson:1985uq}
\bibitem{Dyson:1985uq}
\bibinfo{author}{Freeman~J \surnamestart Dyson\surnameend}
  (\bibinfo{year}{1985}): \bibinfo{title}{{Origins of life}}.
\newblock \bibinfo{publisher}{Cambridge: Cambridge University Press}.

\bibitemdeclare{article}{Eigen1977a}
\bibitem{Eigen1977a}
\bibinfo{author}{M~\surnamestart Eigen\surnameend} \&
  \bibinfo{author}{P~\surnamestart Schuster\surnameend} (\bibinfo{year}{1977}):
  \bibinfo{title}{{The hypercycle. A principle of natural
  self-organization. Part A: Emergence of the hypercycle.}}
\newblock {\sl \bibinfo{journal}{Die Naturwissenschaften}}
  \bibinfo{volume}{64}(\bibinfo{number}{11}), pp. \bibinfo{pages}{541--65}.
\doi{10.1007/BF00450633}
\newblock \urlprefix\url{http://www.ncbi.nlm.nih.gov/pubmed/593400}.

\bibitemdeclare{article}{Eigen1988}
\bibitem{Eigen1988}
\bibinfo{author}{Manfred \surnamestart Eigen\surnameend} \&
  \bibinfo{author}{John \surnamestart Mccaskill\surnameend}
  (\bibinfo{year}{1988}): \bibinfo{title}{{Molecular Quasi-Specie}}.
\newblock {\sl \bibinfo{journal}{J Phys Chem}} \bibinfo{volume}{81}, pp.
\doi{10.1021/j100335a010}
  \bibinfo{pages}{6881--6891}.

\bibitemdeclare{article}{Eigen1978}
\bibitem{Eigen1978}
\bibinfo{author}{Manfred \surnamestart Eigen\surnameend} \&
  \bibinfo{author}{Peter \surnamestart Schuster\surnameend}
  (\bibinfo{year}{1978}): \bibinfo{title}{{The Hypercycle: a Principle of
  Natural Self-Organisation, Part B}}.
\newblock {\sl \bibinfo{journal}{Naturwissenschaften}}
  \bibinfo{volume}{65}(\bibinfo{number}{7}), pp. \bibinfo{pages}{7--41}.
\newblock
\doi{10.1007/BF00439699}
  \urlprefix\url{http://www.springerlink.com/index/R133207N06736808.pdf}.

\bibitemdeclare{article}{ac:ES78c}
\bibitem{ac:ES78c}
\bibinfo{author}{Manfred \surnamestart Eigen\surnameend} \&
  \bibinfo{author}{Peter \surnamestart Schuster\surnameend}
  (\bibinfo{year}{1978}): \bibinfo{title}{{The Hypercycle: a Principle of
  Natural Self-Organisation, Part C}}.
\newblock {\sl \bibinfo{journal}{Naturwissenschaften}}
  \bibinfo{volume}{65}(\bibinfo{number}{7}), pp. \bibinfo{pages}{341--369}.
\doi{10.1007/BF00439699}

\bibitemdeclare{article}{JSC2011}
\bibitem{JSC2011}
\bibinfo{author}{A.~\surnamestart Filisetti\surnameend},
  \bibinfo{author}{A.~\surnamestart Graudenzi\surnameend},
  \bibinfo{author}{R.~\surnamestart Serra\surnameend},
  \bibinfo{author}{M.~\surnamestart Villani\surnameend},
  \bibinfo{author}{D.~\surnamestart De~Lucrezia\surnameend},
  \bibinfo{author}{R.~M. \surnamestart Fuchslin\surnameend},
  \bibinfo{author}{S.~A. \surnamestart Kauffman\surnameend},
  \bibinfo{author}{N.~\surnamestart Packard\surnameend} \&
  \bibinfo{author}{I.~\surnamestart Poli\surnameend} (\bibinfo{year}{2011}):
  \bibinfo{title}{A stochastic model of the emergence of autocatalytic
  cycles}.
\newblock {\sl \bibinfo{journal}{Journal of Systems Chemistry}}
  \bibinfo{volume}{2}(\bibinfo{number}{2}), pp.
  \bibinfo{pages}{doi:10.1186/1759--2208--2--2}.

\bibitemdeclare{inproceedings}{Filisetti2011b}
\bibitem{Filisetti2011b}
\bibinfo{author}{A~\surnamestart Filisetti\surnameend},
  \bibinfo{author}{A~\surnamestart Graudenzi\surnameend},
  \bibinfo{author}{R~\surnamestart Serra\surnameend},
  \bibinfo{author}{M~\surnamestart Villani\surnameend},
  \bibinfo{author}{D~\surnamestart {De Lucrezia}\surnameend} \&
  \bibinfo{author}{I~\surnamestart Poli\surnameend} (\bibinfo{year}{2011}):
  \bibinfo{title}{{The role of energy in a stochastic model of the
  emergence of autocatalytic sets}}.
\newblock In \bibinfo{editor}{Lenaerts \surnamestart T\surnameend},
  \bibinfo{editor}{Giacobini \surnamestart M\surnameend},
  \bibinfo{editor}{H~\surnamestart Bersini\surnameend},
  \bibinfo{editor}{P~\surnamestart Bourgine\surnameend},
  \bibinfo{editor}{M~\surnamestart Dorigo\surnameend} \&
  \bibinfo{editor}{R~\surnamestart Doursat\surnameend}, editors: {\sl
  \bibinfo{booktitle}{Advances in Artificial Life, ECAL 2011 Proceedings of the
  Eleventh European Conference on the Synthesis and Simulation of Living
  Systems}}, \bibinfo{publisher}{MIT Press, Cambridge, MA}, pp.
  \bibinfo{pages}{227--234}.

\bibitemdeclare{article}{TIB2011}
\bibitem{TIB2011}
\bibinfo{author}{A.~\surnamestart Filisetti\surnameend},
  \bibinfo{author}{A.~\surnamestart Graudenzi\surnameend},
  \bibinfo{author}{R.~\surnamestart Serra\surnameend},
  \bibinfo{author}{M.~\surnamestart Villani\surnameend}, \bibinfo{author}{R.~M.
  \surnamestart Fuchslin\surnameend}, \bibinfo{author}{N.~H. \surnamestart
  Packard\surnameend}, \bibinfo{author}{S.A. \surnamestart Kauffman\surnameend}
  \& \bibinfo{author}{I.~\surnamestart Poli\surnameend} (\bibinfo{year}{2011}):
  \bibinfo{title}{A stochastic model of autocatalytic reaction
  networks.}
\newblock {\sl \bibinfo{journal}{Theory in biosciences = Theorie in den
  Biowissenschaften}} \bibinfo{volume}{DOI: 10.1007/s12064-011-0136-x.}

\bibitemdeclare{inproceedings}{Filisetti:2010fk}
\bibitem{Filisetti:2010fk}
\bibinfo{author}{Alessandro \surnamestart Filisetti\surnameend},
  \bibinfo{author}{Roberto \surnamestart Serra\surnameend},
  \bibinfo{author}{Marco \surnamestart Villani\surnameend},
  \bibinfo{author}{Rudolf~M \surnamestart F\"{u}chslin\surnameend},
  \bibinfo{author}{Norman~H \surnamestart Packard\surnameend},
  \bibinfo{author}{Stuart~A \surnamestart Kauffman\surnameend} \&
  \bibinfo{author}{Irene \surnamestart Poli\surnameend} (\bibinfo{year}{2010}):
  \bibinfo{title}{A stochastic model of autocatalytic reaction
  networks}.
\newblock In: {\sl \bibinfo{booktitle}{Proceedings of the European Conference
  on Complex Systems (ECCS)}}, \bibinfo{address}{Lisbon, September 13-17}.

\bibitemdeclare{article}{Gillespie:1977fv}
\bibitem{Gillespie:1977fv}
\bibinfo{author}{Daniel~T \surnamestart Gillespie\surnameend}
  (\bibinfo{year}{1977}): \bibinfo{title}{{Exact Stochastic Simulation of
  Coupled Chemical Reactions}}.
\newblock {\sl \bibinfo{journal}{The Journal of Physical Chemistry}}
  \bibinfo{volume}{81}(\bibinfo{number}{25}), pp. \bibinfo{pages}{2340--2361}.
\doi{10.1021/j100540a008}

\bibitemdeclare{article}{Hordijk2010}
\bibitem{Hordijk2010}
\bibinfo{author}{Wim \surnamestart Hordijk\surnameend}, \bibinfo{author}{Jotun
  \surnamestart Hein\surnameend} \& \bibinfo{author}{Mike \surnamestart
  Steel\surnameend} (\bibinfo{year}{2010}):
  \bibinfo{title}{{Autocatalytic Sets and the Origin of Life}}.
\newblock {\sl \bibinfo{journal}{Entropy}}
  \bibinfo{volume}{12}(\bibinfo{number}{7}), pp. \bibinfo{pages}{1733--1742},
  \doi{10.3390/e12071733}.
\newblock \urlprefix\url{http://www.mdpi.com/1099-4300/12/7/1733/}.

\bibitemdeclare{article}{Hordijk:2004vl}
\bibitem{Hordijk:2004vl}
\bibinfo{author}{Wim \surnamestart Hordijk\surnameend} \& \bibinfo{author}{Mike
  \surnamestart Steel\surnameend} (\bibinfo{year}{2004}):
  \bibinfo{title}{{Detecting autocatalytic, self-sustaining sets in
  chemical reaction systems.}}
\newblock {\sl \bibinfo{journal}{Journal of theoretical biology}}
  \bibinfo{volume}{227}(\bibinfo{number}{4}), pp. \bibinfo{pages}{451--461},
  \doi{10.1016/j.jtbi.2003.11.020}.
\newblock \urlprefix\url{http://www.ncbi.nlm.nih.gov/pubmed/15038982}.

\bibitemdeclare{article}{Jain:1998fk}
\bibitem{Jain:1998fk}
\bibinfo{author}{Sanjai \surnamestart Jain\surnameend} \&
  \bibinfo{author}{Sandeep \surnamestart Krishna\surnameend}
  (\bibinfo{year}{1998}): \bibinfo{title}{{Autocatalytic set and the
  growth of complexity in an evolutionary model}}.
\newblock {\sl \bibinfo{journal}{Phys Rev Lett}} \bibinfo{volume}{81}, pp.
  \bibinfo{pages}{5684--5687}.
\doi{10.1103/PhysRevLett.81.5684}

\bibitemdeclare{book}{Kaneko:2006eu}
\bibitem{Kaneko:2006eu}
\bibinfo{author}{Kunihiko \surnamestart Kaneko\surnameend}
  (\bibinfo{year}{2006}): \bibinfo{title}{{Life: An Introduction to
  Complex Systems Biology (Understanding Complex Systems)}}.
\newblock \bibinfo{publisher}{Springer-Verlag New York, Inc.},
  \bibinfo{address}{Secaucus, NJ, USA}.

\bibitemdeclare{article}{ensemble}
\bibitem{ensemble}
\bibinfo{author}{S.~\surnamestart Kauffman\surnameend} (\bibinfo{year}{2004}):
  \bibinfo{title}{{A proposal for using the ensemble approach to
  understand genetic regulatory networks}}.
\newblock {\sl \bibinfo{journal}{Journal of Theoretical Biology}}
  \bibinfo{volume}{230}(\bibinfo{number}{4}), pp. \bibinfo{pages}{581--590}.
\doi{10.1016/j.jtbi.2003.12.017}

\bibitemdeclare{article}{Kauffman:1986mi}
\bibitem{Kauffman:1986mi}
\bibinfo{author}{S~A \surnamestart Kauffman\surnameend} (\bibinfo{year}{1986}):
  \bibinfo{title}{{Autocatalytic sets of proteins.}}
\newblock {\sl \bibinfo{journal}{J Theor Biol}}
  \bibinfo{volume}{119}(\bibinfo{number}{1}), pp. \bibinfo{pages}{1--24}.
\doi{10.1016/S0022-5193(86)80047-9}

\bibitemdeclare{article}{Mossel2005}
\bibitem{Mossel2005}
\bibinfo{author}{Elchanan \surnamestart Mossel\surnameend} \&
  \bibinfo{author}{Mike \surnamestart Steel\surnameend} (\bibinfo{year}{2005}):
  \bibinfo{title}{{Random biochemical networks: the probability of
  self-sustaining autocatalysis.}}
\newblock {\sl \bibinfo{journal}{Journal of theoretical biology}}
  \bibinfo{volume}{233}(\bibinfo{number}{3}), pp. \bibinfo{pages}{327--36},
  \doi{10.1016/j.jtbi.2004.10.011}.

\bibitemdeclare{article}{Segre2000}
\bibitem{Segre2000}
\bibinfo{author}{D~\surnamestart Segr\'{e}\surnameend} \&
  \bibinfo{author}{D~\surnamestart Lancet\surnameend} (\bibinfo{year}{2000}):
  \bibinfo{title}{{Composing life}}.
\newblock {\sl \bibinfo{journal}{EMBO reports}}
  \bibinfo{volume}{1}(\bibinfo{number}{3}), pp. \bibinfo{pages}{217--22},
  \doi{10.1093/embo-reports/kvd063}.
\newblock
  \urlprefix\url{http://www.nature.com/embor/journal/v1/n3/full/embor574.html}.

\bibitemdeclare{article}{Segre:1998sf}
\bibitem{Segre:1998sf}
\bibinfo{author}{D~\surnamestart Segre\surnameend},
  \bibinfo{author}{D~\surnamestart Lancet\surnameend},
  \bibinfo{author}{O~\surnamestart Kedem\surnameend} \&
  \bibinfo{author}{Y~\surnamestart Pilpel\surnameend} (\bibinfo{year}{1998}):
  \bibinfo{title}{Graded Autocatalysis Replication Domain (GARD):\\
  kinetic analysis of self-replication in mutually catalytic sets.}
\newblock {\sl \bibinfo{journal}{Orig Life Evol Biosph}}
  \bibinfo{volume}{28}(\bibinfo{number}{4-6}), pp. \bibinfo{pages}{501--514}.
\doi{10.1023/A:1006583712886}

\end{thebibliography}
\end{document}